# Gravity Control produced by a Thermoionic Current through the Air at Very Low Pressure

## Fran De Aquino

Maranhao State University, Physics Department, S.Luis/MA, Brazil.

It was observed that samples hung above a thermoionic current exhibit a weight decrease directly proportional to the intensity of the current. The observed phenomenon appears to be absolutely new and unprecedented in the literature and can not be understood in the framework of the general relativity. It is pointed out the possibility that this unexpected effect is connected with a possible correlation between gravity and electromagnetism.

## I. INTRODUCTION

It will be described an experiment in which there has been observed a *strong* decrease in the weight of samples hung above a thermoionic current produced inside a vacuum chamber. The percentage of weight decrease is the same for samples of different masses and chemical compositions. The effect does not seem to diminish with increases in elevation above the current.

This unexpected phenomenon appears to be unprecedented in the literature. On the other hand, the experiment is simple of being performed, and can be easily replicated.

## II. EXPERIMENTAL

### 1. General description of the experimental set-up.

A *Fe* sample (iron tube with 1.103kg; outer diameter = 100mm; inner diameter = 80mm; height = 50mm) was placed inside a vacuum chamber (at 0.001militorr) and above the oscillating thermoionic current $(f = 60Hz)$ produced by the electrons flow from the cathode to plate as shown in Fig.1. The distance between the cathode and the plate is

10mm (exactly the thickness of the iron tube above the current). The area of the plate is

$$S_{plate} = 2\pi \; bl =$$
$$= 2\pi(0.050m)(0.030m) = 9.42 \times 10^{-3} \, m^2$$

The area of the cathode is

$$S_{cathode} = 2\pi \; al =$$
$$= 2\pi(0.040m)(0.030m) = 7.56 \times 10^{-3} \, m^2$$

Therefore the average area is $S = 8.48 \times 10^{-3} \, m^2$.

The cathode, the filament and the plate are made of *tungsten*.

Two piezoelectric sensors for force (S1,S2) are placed as shown in Fig.1, in order to measure the weight of the sample (tube) during the flow of thermoionic current. When the thermoionic current is null the sensor S1 shows the normal weight of the sample. Note that there is a central disk inside the tube to support the sample. Above this disk (very close to the disk) it is the sensor S2. The function of the sensor S2 is to measure the weight of the sample, in case this weight becomes negative (inversion).

### 2. Conductivity Measurements.

Atmospheric air conductivity is proportional to both the atmospheric ion concentration $\eta$, and the average mobility $\mu$ of the air ion population. Molecular ions with μ > 0.5 cm2V-1s-1 are conventionally defined as "small ions" [1]. The unipolar air conductivity $\sigma_{air}$ can be written as

$$\sigma_{air} = e \int_{0.5cm^2V^{-1}s^{-1}}^{\infty} \eta d\mu \qquad (1)$$

where $e$ is the charge of the electron. Typical surface values of atmospheric air conductivity are $2-100 \times 10^{-15}\, S.m^{-1}$ [2,3].

The air conductivity between the cathode and the plate is much greater than these values due to the large concentration of free electrons. It can be evaluated by comparing the Gauss' law $(\phi_E = ES = q/\varepsilon_0)$ with $J = \sigma E$, i.e.,

$$J = \frac{i}{S} = \sigma \frac{q}{S\varepsilon_0} = \sigma \frac{(CV)}{S\varepsilon_0} \qquad (2)$$

hence

$$\sigma_{air} = \frac{\varepsilon_0 i}{CV} \qquad (3)$$

where $V$ is the voltage from the anode (plate) to cathode and $C$ is the capacitance of the cylindrical capacitor (cathode/plate) given by

$$C = \frac{2\pi\varepsilon_0 l}{ln(b/a)} \qquad (4)$$

where $a$ and $b$ are the radii of the outer and central electrodes; $l$ is the height of the electrodes(cathode and plate).

Since the Langmuir-Child law states that the *thermoionic current* density is given by

$$J = \frac{4}{9}\varepsilon_0 \sqrt{\frac{2e}{m_e}} \frac{V^{\frac{3}{2}}}{d^2} = \alpha \frac{V^{\frac{3}{2}}}{d^2} = 2.33 \times 10^{-6} \frac{V^{\frac{3}{2}}}{d^2} \qquad (5)$$

where $\alpha$ is the called *Child's constant*.

Then Eq.(3) gives

$$\sigma_{air} = \frac{\varepsilon_0}{CV}(JS) = \left[\frac{\alpha S\, ln(b/a)}{2\pi\, d^2 l}\right] V^{\frac{1}{2}} \qquad (6)$$

As shown in Fig.1, $b = 50mm$ ; $a = 40mm$ and $l = 30mm$; $S = 8.48 \times 10^{-3}\, m^2$. Thus Eq. (6) gives

$$\sigma_{air} = 2.34 \times 10^{-4} V^{\frac{1}{2}} \qquad (7)$$

### 3. Pressure, Temperature and Mass Density of the air.

The chamber was sealed and evacuated to about 0.001 millitorr using a vacuum pump system (TP-70-2DR oil-free system; wide range vacuum gauge: atmosphere to 10$^{-8}$torr).

The cathode temperature reach ~2000K at this temperature and 0.001 millitorr, the density of the air between the cathode and plate is

$$\rho_{air} \cong 3.1 \times 10^{-10} Kg.m^{-3} \qquad (8)$$

## III. RESULTS

We have started with voltage $V_{rms} = 100V$ at $60Hz$. Next, the voltage was progressively increased to $200V$, $300V$, $400V$ and $500V$. Table1 presents the weight behavior of the sample measured by the sensors S1 and S2.

## IV. DISCUSSION

The weight behavior of our samples shows strong variations, which apparently can be explained as due to the *thermoionic current*, which flows from the cathode to plate. When this current is removed the effects disappear.

In a previous work [4] we have shown that when an alternating electric current passes through a substance its *gravitational mass* is reduced in accordance with the following expression

$$m_g = \left\{ 1 - 2\left[ \sqrt{1 + \left(i^4 \mu / 64\pi^3 c^2 \rho^2 S^4 f^3 \sigma\right)} - 1 \right] \right\} m_i \quad (9)$$

In this equation $i$ refers to the instantaneous electric current[†] ; $\mu = \mu_r \mu_0$ is the magnetic permeability of the substance; $c$ is the speed of light; $\rho$ , $S$ and $\sigma$ are respectively the density (kg/m³), the area of the cross section (m²) and the electric conductivity (S/m) of the substance; $m_i$ is the *inertial mass* and $f$ the frequency of the electric current (Hz).

It was also shown that there is an additional effect of *gravitational shielding* produced by the substance under these conditions. Above the substance the gravity acceleration $g'$ is reduced at the same ratio $\chi = m_g / m_i$ , i.e., $g' = \chi g$ .

If the substance is the *air at very low pressure* such as the air between the cathode and the plate of vacuum chamber presented in Fig.1, for example with the following characteristics: relative *magnetic permeability=* $\mu_r \cong 1$; electrical *conductivity* $= \sigma_{air} = 2.34 \times 10^{-4} V^{\frac{1}{2}}$; *density=* $\rho_{air} \cong 3.1 \times 10^{-10} kg m^{-3}$ at $0.001$ millitorr and ~2000K. Then for $f = 60 Hz$ and (9) gives

_______________________

[†] It is known that the current is a spectrum in the *Fourier* domain. Here the spectrum has a D.C component, a 60Hz component, a 120Hz component, a 180Hz component, etc. The D.C component does not affect $m_g$ since the effect just occurs for oscillating currents [4]. On the other hand, the effect of the 120Hz component is negligible (in voltage range investigated) in respect to the 60Hz component since the factor $\left(i^4 \mu / 64\pi^3 c^2 \rho^2 S^4 f^3 \sigma\right)$ in Eq. (9) is $(120/60)^3 = 8$ times smaller than in the case of the 60Hz component. For the 180Hz component the factor is $(180/60)^3 = 27$ times smaller than in the case of the 60Hz component. Therefore here we will consider the 60Hz component only.

$$m_{g(air)} = \left\{ 1 - 2\left[ \sqrt{1 + 4.26 \times 10^{-16} V_{max}^{5.5}} - 1 \right] \right\} m_{i(air)} \quad (10)$$

Therefore, due to the *gravitational shielding effect* produced by the decreasing of $m_{g(air)}$, the gravity acceleration *above* the air between the cathode and the plate will be given by

$$g' = \chi g = \frac{m_{g(air)}}{m_{i(air)}} g =$$
$$= \left\{ 1 - 2\left[ \sqrt{1 + 4.26 \times 10^{-16} V_{max}^{5.5}} - 1 \right] \right\} g$$

Consequently, the weight of the sample, $P_s$ , will be given by

$$P_s = m_{gs} g' = m_{is} g' \qquad (11)$$

In the case of the Fe sample: $m_{is} = 1.103 Kg$ . The weight of the sample, measured during the increase of voltage is presented on Table 1. The theoretical values, calculated by means of (10), are also on Table 1 to be compared with those supplied by the experiment.

Afterwards, the iron tube was replaced by similar tubes (same dimensions) with different masses and chemical compositions. The values of the gravity acceleration $g'$ for these samples were the same.

## V. CONSEQUENCES

The experimental results point to the possibility of conversion of gravitational energy into mechanical energy and electrical energy. Consider for example the system presented in Fig.3 (a). Basically it is a motor with massive iron rotor and a gravity control cell. This cell is similar to the diode presented in Fig.1 (see Fig.3(b)), it is placed below the rotor in order to become *negative* the acceleration of gravity inside *half* of the rotor $\left(g' = -ng\right)$, as showed in Fig.3 (a). Obviously this causes a torque $T = (-F' + F)r$ and the rotor spins with angular velocity $\omega$. The average power, $P$, of the motor is given by

$$P = T\omega = [(-F' + F)r]\omega \qquad (12)$$

where

$$F' = \tfrac{1}{2}m_g g' \qquad F = \tfrac{1}{2}m_g g$$

and $m_g \cong m_i$ (mass of the rotor). Thus, Eq. (12) gives

$$P = (n+1)\frac{m_i g \omega \ r}{2} \qquad (13)$$

On the other hand, we have that

$$-g' + g = \omega^2 r \qquad (14)$$

Therefore the angular speed of the rotor is given by

$$\omega = \sqrt{\frac{(n+1)g}{r}} \qquad (15)$$

By substituting (15) into (13) we obtain the expression of the average power of the *gravitational motor*, i.e.,

$$P = \tfrac{1}{2}m_i \sqrt{(n+1)^3 g^3 r} \qquad (16)$$

Now consider an electric generator coupling to the gravitational motor in order to produce electric energy. Since $\omega = 2\pi f$ then for $f = 60Hz$ we have $\omega = 120\pi rad.s^{-1} = 3600 \ rpm$. Therefore for $\omega = 120\pi rad.s^{-1}$ and $n = 788$ $(V_{max} \cong 5.5KV)^{\ddagger}$ the Eq. (15) tell us that we must have

---

$\ddagger$ Note that, according to Eq.(9), this voltage can be strongly reduced by decreasing the density of the air, $\rho$, inside the gravity control cell.

$$r = \frac{(n+1)g}{\omega^2} = 0.0545m$$

Since $r = R/3$ and $m_i = \rho\pi R^2 h$ where $\rho$, $R$ and $h$ are respectively the mass density, the radius and the height of the rotor then for $h = 1.0m$ and $\rho = 7800Kg.m^{-3}$ (iron) we obtain

$$m_i = 654.1kg$$

Then Eq. (16) gives

$$P \cong 5.2 \times 10^7 \ watts \cong 52MW \qquad (17)$$

This shows that the gravitational motor can be used to yield electric energy at large scale.

## VI. CONCLUSION

In the frequency investigated (60Hz), the weight of the samples decrease strongly with the increase of the intensity of the thermoionic current, and, as we have shown, the weight of the samples can even become *negative* (inversion).

The experimental observations described in this work are absolutely new and unprecedented. They tell us about a part of Gravitation Theory which is unknown.

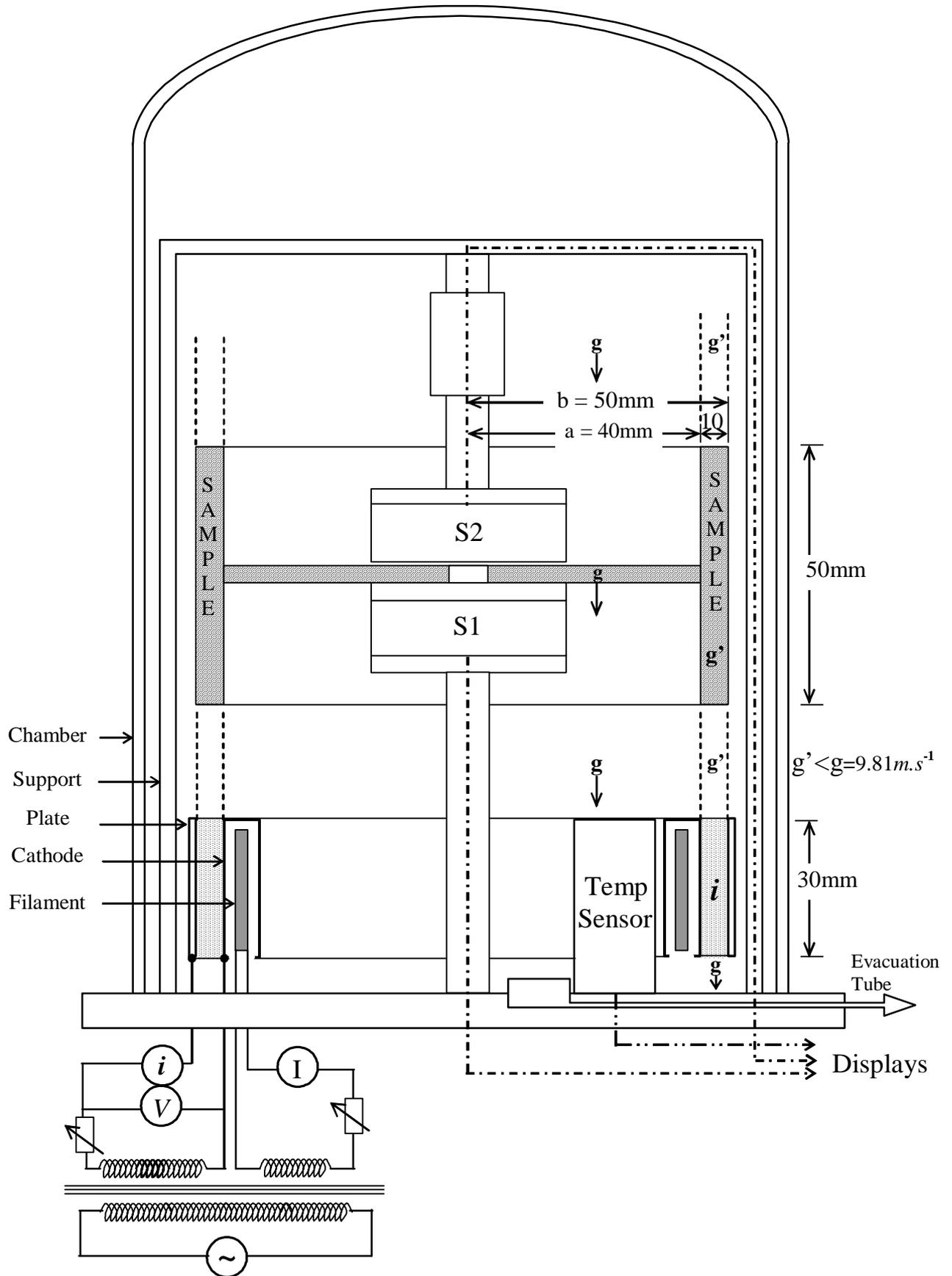

Fig.1- Experimental set-up.

| $V_{rms}$ (V) | $V_{max}$ (V) | Weight | | $g'/g$ | | Sensor |
|---|---|---|---|---|---|---|
| | | Exp (N) | Theo (N) | Exp | Theo | |
| 0 | 0 | 10.82 | 10.82 | 1 | 1 | |
| 100 | 141.42 | 10.60 | 10.71 | 0.98 | 0.99 | |
| 200 | 282.84 | 10.49 | 10.60 | 0.97 | 0.98 | S1 |
| 300 | 424.26 | 9.30 | 9.52 | 0.86 | 0.88 | |
| 400 | 565.68 | 4.44 | 5.19 | 0.41 | 0.48 | |
| 500 | 707.11 | -7.03 | -4.98 | -0.65 | -0.46 | S2 |

Table 1 - Influence of *thermoionic current* (60Hz) on the weight of the sample. Experimental data are the average of 10 measurements. The standard deviation of the single data is between 3 and 5%.

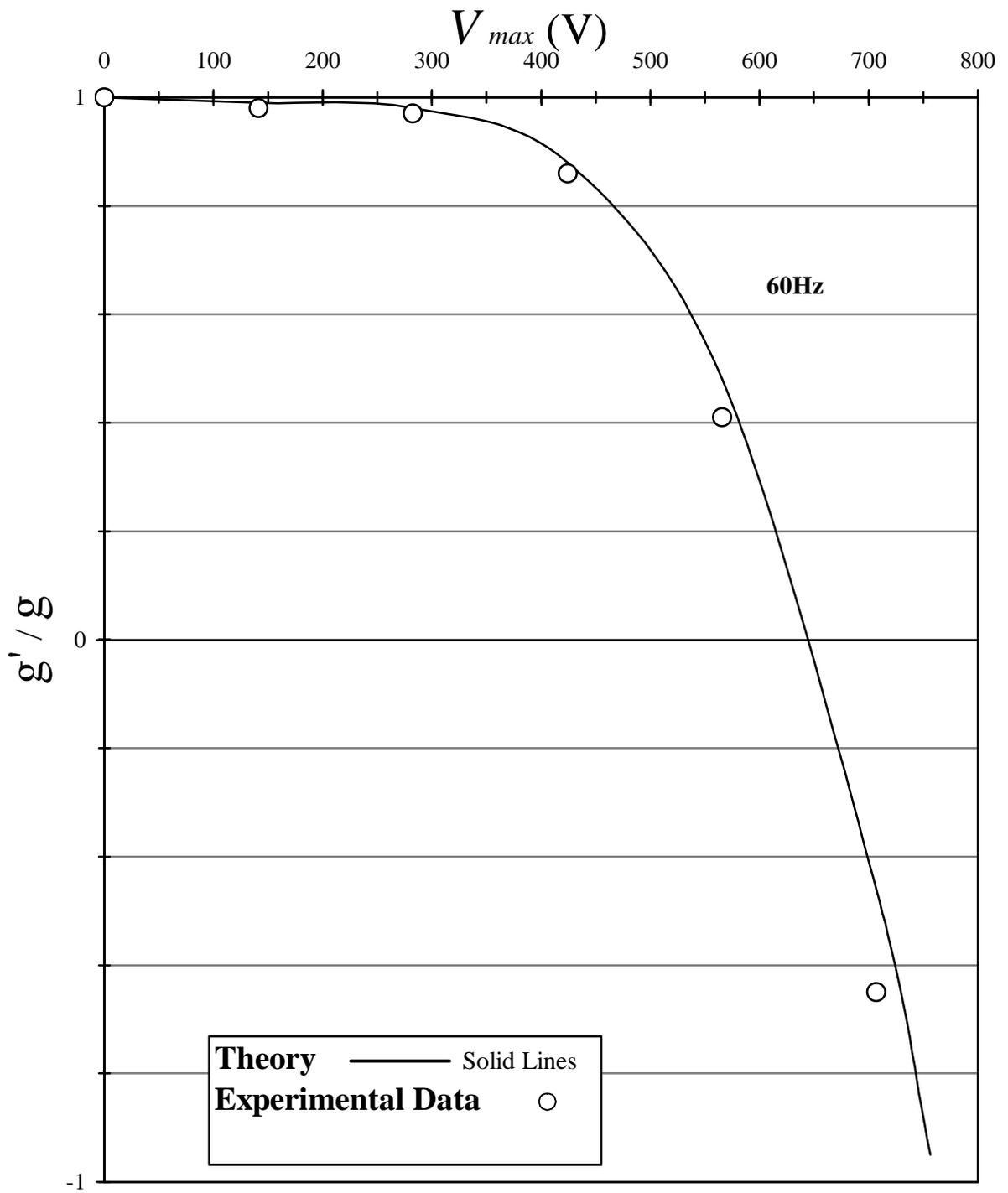

**Fig. 2**- Distribution of the correlation    g'/g  as a function of   $V$ max.

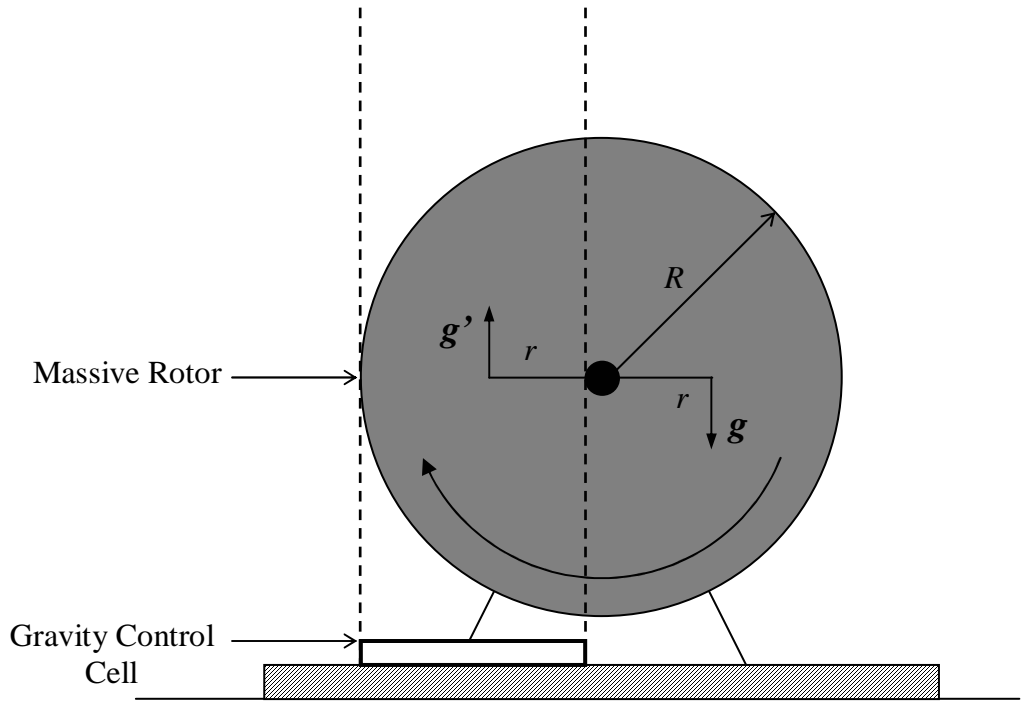

(a)

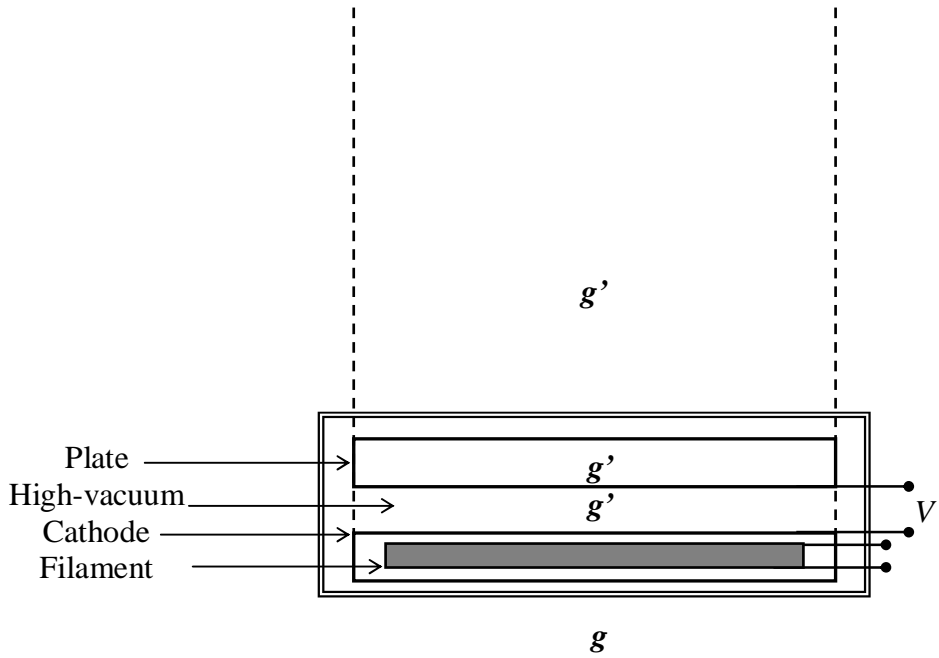

Gravity Control Cell
(b)

**Fig. 3** – The Gravitational Motor